\documentclass{aastex631}
\usepackage{graphicx}
\usepackage{natbib}

\shorttitle{filament formation observed by CHASE}
\shortauthors{Li et al.}

\begin{document}

\title{Formation of a long filament through the connection of two filament segments observed by CHASE}
 

\author[0000-0001-6024-8399]{H. T. Li}
\affiliation{School of Astronomy and Space Science, Nanjing University, Nanjing 210023, People$'$s Republic of China; xincheng@nju.edu.cn}
\affiliation{Key Laboratory of Modern Astronomy and Astrophysics (Nanjing University), Ministry of Education, Nanjing 210023, People$'$s Republic of China}

\author[0000-0003-2837-7136]{X. Cheng}
\affiliation{School of Astronomy and Space Science, Nanjing University, Nanjing 210023, People$'$s Republic of China; xincheng@nju.edu.cn}
\affiliation{Key Laboratory of Modern Astronomy and Astrophysics (Nanjing University), Ministry of Education, Nanjing 210023, People$'$s Republic of China}

\author[0000-0002-9908-291X]{Y. W. Ni}
\affiliation{School of Astronomy and Space Science, Nanjing University, Nanjing 210023, People$'$s Republic of China; xincheng@nju.edu.cn}
\affiliation{Key Laboratory of Modern Astronomy and Astrophysics (Nanjing University), Ministry of Education, Nanjing 210023, People$'$s Republic of China}

\author[0000-0001-7693-4908]{C. Li}
\affiliation{School of Astronomy and Space Science, Nanjing University, Nanjing 210023, People$'$s Republic of China; xincheng@nju.edu.cn}
\affiliation{Key Laboratory of Modern Astronomy and Astrophysics (Nanjing University), Ministry of Education, Nanjing 210023, People$'$s Republic of China}

\author[0000-0002-7544-6926]{S. H. Rao}
\affiliation{School of Astronomy and Space Science, Nanjing University, Nanjing 210023, People$'$s Republic of China; xincheng@nju.edu.cn}
\affiliation{Key Laboratory of Modern Astronomy and Astrophysics (Nanjing University), Ministry of Education, Nanjing 210023, People$'$s Republic of China}

\author[0000-0002-4205-5566]{J. H. Guo}
\affiliation{School of Astronomy and Space Science, Nanjing University, Nanjing 210023, People$'$s Republic of China; xincheng@nju.edu.cn}
\affiliation{Key Laboratory of Modern Astronomy and Astrophysics (Nanjing University), Ministry of Education, Nanjing 210023, People$'$s Republic of China}
\affiliation{Centre for Mathematical Plasma Astrophysics, Department of Mathematics, KU Leuven, Celestijnenlaan 200B,B-3001 Leuven, Belgium}

\author[0000-0002-4978-4972]{M. D. Ding}
\affiliation{School of Astronomy and Space Science, Nanjing University, Nanjing 210023, People$'$s Republic of China; xincheng@nju.edu.cn}
\affiliation{Key Laboratory of Modern Astronomy and Astrophysics (Nanjing University), Ministry of Education, Nanjing 210023, People$'$s Republic of China}

\author[0000-0002-7289-642X]{P. F. Chen}
\affiliation{School of Astronomy and Space Science, Nanjing University, Nanjing 210023, People$'$s Republic of China; xincheng@nju.edu.cn}
\affiliation{Key Laboratory of Modern Astronomy and Astrophysics (Nanjing University), Ministry of Education, Nanjing 210023, People$'$s Republic of China}


\begin{abstract}
We present imaging and spectroscopic diagnostics of a long filament during its formation with the observations from the Chinese H$\alpha$ Solar Explorer and Solar Dynamics Observatory. The seed filament first appeared at about 05:00 UT on 2022 September 13. Afterwards, it grew gradually and connected to another filament segment nearby, building up a long filament at about 20:00 UT on the same day. The CHASE H$\alpha$ spectra show an obvious centroid absorption with mild broadening at the main spine of the long filament, which is interpreted as the evidence of filament material accumulation. More interestingly, near the footpoints of the filament, persistent redshifts have been detected in the H$\alpha$ spectra during the filament formation, indicating continuous drainage of filament materials. Furthermore, through inspecting the extreme ultraviolet images and magnetograms, it is found that EUV jets and brightenings appeared repeatedly at the junction of the two filament segments, where opposite magnetic polarities converged and canceled to each other continuously. These results suggest the occurrence of intermittent magnetic reconnection that not only connects magnetic structures of the two filament segments but also supplies cold materials for the filament channel likely by the condensation of injected hot plasma, even though a part of cold materials fall down to the filament footpoints at the same time.
\end{abstract}


\section{Introduction} \label{sec:intro}
Solar filaments are cold and dense plasma suspended in the hot and tenuous corona above the polarity inversion lines (PILs) of magnetic field at the photosphere \citep{Tandberg.1974.Book, Martin.1998.Solphys,Chenpengfei.2020.RAA}. They often appear as dark structures on the solar disk and are associated with an absorption in the H$\alpha$ line. When appearing above the solar limb, they are also called prominences with brighter emissions than the background \citep{Mackay.2010.SSR}. 

Studies about solar filaments can be roughly divided into two categories. The first category is focusing on the nature and formation mechanisms of the filament magnetic structure, which is also named filament channel \citep{Schmieder.2014.AA, Patsourakos.2020.SSR}. At present, it is widely accepted that the magnetic structure of filaments could be either sheared arcades \citep{Antiochos.1994.ApJL} or twisted magnetic flux ropes \citep{Aulanier.1998.AA,Chengxin.2014.ApJL, Chengxin.2015.ApJ}, as both of them contain concave magnetic dips that are able to support filament materials against the gravity. 

In the past decades, more attention was paid to the formation of filament channels consisting of a twisted flux rope, possibly because about 89\% of filaments are supported by flux ropes \cite{Ouyangyu.2017.ApJ}. The twisted filament channel can be gradually formed through photospheric rotational motions that continuously inject twists into the channel \citep{Torok.2003.AA, Yanxiaoli.2015.ApJS, Yanxiaoli.2016.ApJ}. Besides, magnetic reconnection is argued to be the more important mechanism because it can directly build up a coherent flux rope by reconnecting sheared arcades \citep{vanBallegooijen.1989.ApJ, Moore.2001.ApJ,Torok.2011.ApJ}. This was well supported by observations where a long filament (hot channel) was formed by the reconnection of two filament segments (sheared arcades) \citep{Schmieder.2004.Solphys, Joshi.2014.ApJ, Chengxin.2015.ApJ, Zhuchunming.2015.ApJ, Chenhuadong.2016.ApJL}. In addition, erupting filaments were also observed occasionally to interact with a nearby filament, giving rise to a distinct filament with a new footpoint \citep{Jiangyunchun.2014.ApJ, Joshi.2016.ApJ, Yangliheng.2017.ApJ, Fangyue.2023.ApJ}

The second category of studies is on the fine structures of filaments with particular interest in the origin of the filament materials \citep{Zhouyuhao.2020.NA, Wangjincheng.2022.AA}. One possibility is that the cold materials within a filament come from the emergence of a magnetic flux rope containing cold plasma in the lower atmosphere \citep[e.g.,][]{Manchester.2004.ApJ,Okamoto2008ApJ,Cheung2014LRSP,Zhaoxiaozhou.2017.ApJ}. However, it should be pointed out that the entire emergence of a twisted flux rope from below the photosphere to the corona is difficult \citep{Manchester.2004.ApJ}. Besides flux rope emergence, the cold materials in the chromosphere can be directly transported to the filament channel by surges/jets in a manner of direct injection \citep{Wangyuming.1999.ApJL, Wangjincheng.2018.ApJ}. When the injected plasmas are hot, they need to experience a condensation process, i.e., so-called evaporation-condensation model \citep{Antiochos.1999.ApJ, Xiachun.2012.ApJL, Xiachun.2016.ApJ}. Given that the main difference between the injection and evaporation-condensation models is the temperature of injected plasma, \cite{Huangchujie.2021.ApJL} performed MHD simulations and unified the two models in the same framework. They proposed that if heating takes place at the lower chromosphere, the enhanced pressure will push the cold plasma in the upper chromosphere to the corona through the direct injection way. In contrast, if heating is in the upper chromosphere, the chromospheric plasma will be first heated and evaporated into the corona and then undergoes a condensation process to cold filaments. 

In this paper, we perform a detailed investigation of the formation of a long filament through the interaction of two filament segments with a combination of imaging and spectroscopic observations for the first time. The interesting finding is the coexistence of material accumulation along the filament spine and material drainage toward the filament footpoints, indicating a highly dynamic circulation of materials during the filament formation. In the following, the instruments and the observations are introduced in Section \ref{sec:Ins}, the main results are shown in Section \ref{sec:result}, which is followed by a summary and discussions in Section \ref{sec: sum}.

\section{Instruments and Observations}\label{sec:Ins}
The data we use are mainly from the H$\alpha$ Imaging Spectrograph (HIS; \citealt{Liuqiang.2022.SciCH}) on board the Chinese H$\alpha$ Solar Explorer (CHASE; \citealt{Lichuan.2019.RAA, Lichuan.2022.SciCH}), which was designed to acquire the spectroscopic information of the solar lower atmosphere. It contains two observational modes\footnote{https://ssdc.nju.edu.cn/}: raster scanning mode (RSM) and continuum imaging mode (CIM). The former provides spectra at two wavebands of H$\alpha$ ($6562.8 \pm 3.1$ {\AA}) and \ion{Fe}{1} ($6569.2 \pm 1.4$ {\AA}) with a spectral resolution of 0.024 {\AA} per pixel. The reconstructed monochromatic images of the full solar disk have a spatial resolution of $\sim$ 0.52$''$ per pixel and time cadence of $\sim$ 60 s at all wavelengths. The CIM yields photospheric images at 6689 {\AA} with a full width at half maximum of 13.4 {\AA}. 
       
We also use the data from the Atmospheric Imaging Assembly (AIA; \citealt{Lemen.2012.Solphys}) on board Solar Dynamics Observatory (SDO; \citealt{Pesnell.2012.Solphys}), which provides 7 extreme ultraviolet (EUV) images of the full solar disk with a pixel size of 0.6$''$ and time cadence of 12 s. Simultaneously, the Helioseismic and Magnetic Imager (HMI; \citealp{Schou.2012.Solphys, Scherrer.2012.Solphys}) on board SDO provides the line-of-sight (LOS) magnetograms with a pixel size of $\sim$ 0.6$''$ and time cadence of $\sim$ 45 s, which are used for investigating the magnetic property during the filament formation. Note that, the alignment between the CHASE and SDO images is done by comparing their white-light images, and all images have been reprojected to the same reference time (11:48:10 UT) for a better comparison.

\begin{figure*}[ht!]
	\centering 
	\includegraphics[width=0.9\textwidth,height=0.72\textwidth]{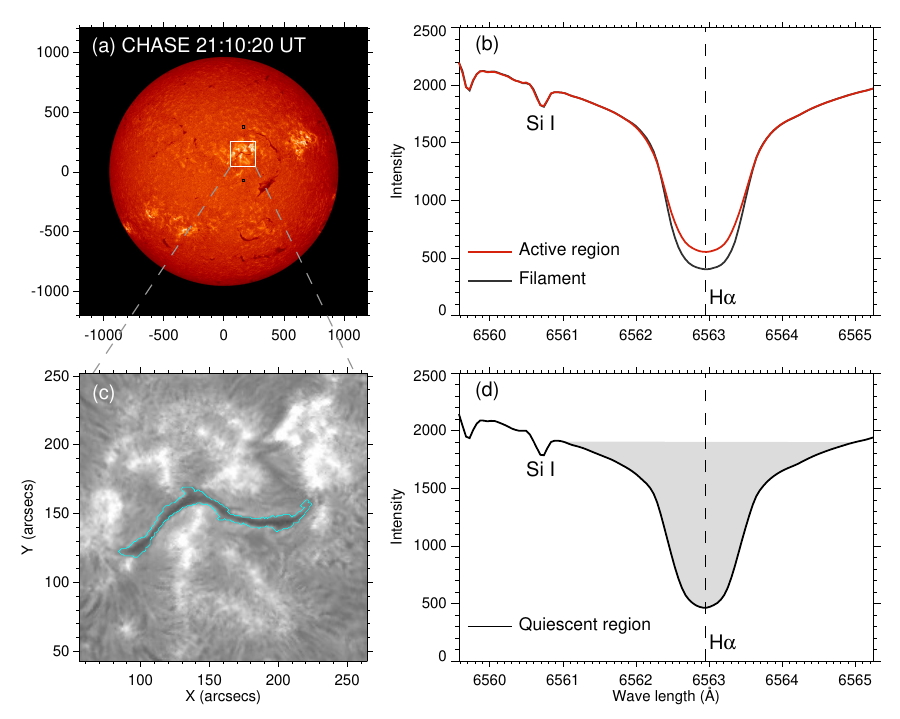}
	\caption{(a) CHASE H$\alpha$ line center image of the full disk showing the location of the target filament pointed out by the white box. The two black boxes show the nearby quiescent regions. (b) The average H$\alpha$ profiles of the entire active region (red) and filament (grey) as indicated by the cyan curve in in panel (c). (d) The H$\alpha$ profile of the quiescent regions that is used for calibration. The shadow region indicates the waveband where the H$\alpha$ line centre is calculated.  \label{fig1}}
\end{figure*}

On 2022 September 13, the CHASE observed the Sun for 15 orbits, each of which lasted approximately 25 minutes. Figure \ref{fig1}a shows the H$\alpha$ line center image of the full Sun observed by CHASE at 21:10:20 UT. The target filament is located near the disk center as indicated by the white box. Figure \ref{fig1}b displays the average H$\alpha$ spectra of the entire target region and filament, which are derived through integrating all pixels within the white box and the filament, respectively (as shown in Figure \ref{fig1}c). Following the method in \citet{Qiuye.2022.SciCH}, the H$\alpha$ line center for all spectra is calibrated assuming that the average spectrum of nearby quiescent regions (as shown in Figure \ref{fig1}d) has a zero Doppler shift. Note that, we take advantage of the waveband of 6561--6565 {\AA} (as shown by the shaded area) for calculating the line centroid, the purpose of which is to avoid the contamination by the \ion{Si}{1} line (6560.58 {\AA}; \citealt{Hongjie.2022.AA}).

\section{Results} \label{sec:result}
\subsection{Filament Formation and Mass accumulation} \label{subsec: sub1}
\begin{figure*}[ht!]
	\centering
	\includegraphics[width=0.9\textwidth,height=0.72\textwidth]{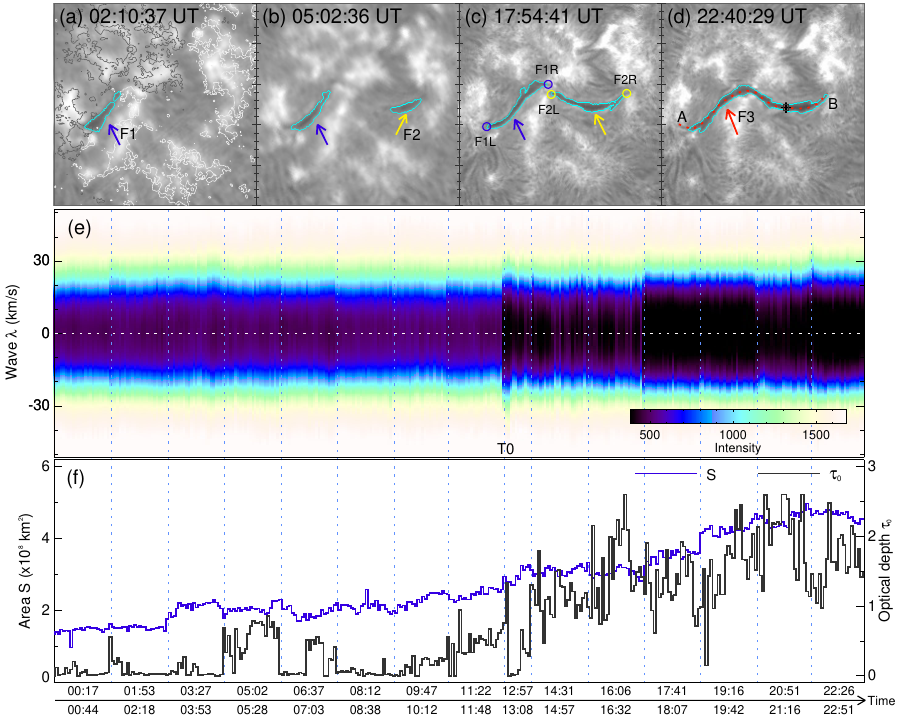}
	\caption{(a)--(d) H$\alpha$ line center images showing the formation and mass accumulation of the long filament. The blue, yellow and red arrows represent the filament F1, F2 and F3, respectively. The contours in cyan indicate the filament region. The white and black contours in panel (a) represent the positive and negative magnetic polarities of $\pm$50G. The blue/yellow circles in pannel (c) mark the footpoints of F1/F2. The plus in panel (d) indicates the position to make the spectrum-time diagram in panel (e). The dashed line in red indicates a slit AB along the filament channel to make the spectrum-distance diagram in Figure \ref{fig3}. (e) H$\alpha$ spectrum-time diagram for the filament spine with the horizontal dashed line showing the line center. (f) Temporal evolutions of the filament area and optical depth. The vertical dashed blue lines in panels (e)--(f) represent the time gaps between the two adjacent orbits, which are hidden here. The start and end times of each orbit are pointed out at above and below of the timeline, respectively. The animation that starts at 00:17 UT and ends at 22:51 UT is available online to show the formation of the long filament F3 with a duration of 28 s. \label{fig2}}
\end{figure*}

The formation and evolution process of the long filament was completely captured by CHASE, as shown in Figures \ref{fig2}a--\ref{fig2}d. From a time sequence of H$\alpha$ line center images, it is found that the seed filament (F1) first appeared as a slightly elongated dark structure at 02:00 UT (pointed out by the blue arrows), lying over the PIL of the active region 13099. At about 05:00 UT, the second seed filament appeared above the PIL of the adjacent active region 13096, as represented by the yellow arrows. Afterwards, with time elapsing, F1 extended toward the northeast and F2 toward the northwest. At 18:00 UT, the right endpoint of F1 merged with the left endpoint of F2, forming an elongated $\Omega$-shaped long filament F3, as indicated by the red arrow. More details can be found in the attached animation of Figure \ref{fig2}.

During the formation of F3, an obvious feature is that more and more filament materials were accumulated in the filament channel, which is indicated by the fact that F3 became darker and darker. To quantify the accumulation of filament materials, we plot the evolution of the H$\alpha$ spectrum of a selected point on the filament spine in Figure \ref{fig2}e, where the point is marked as a plus sign in Figure \ref{fig2}d. It is found that the intensity at the central part of the H$\alpha$ line gradually decreased with the formation of F3, in particular after T0 (11:48 UT). At the same time, the line width increased slightly. According to the cloud model \citep{Beckers.1964.PhDT}, the intensity of a filament can be solved by the radiative transfer equation as: $I=I_0\text{exp}(-\tau)+S[1-\text{exp}(-\tau)]$, where $I_0$ is the intensity of the background quiescent region, $S$ and $\tau$ denote the source function and optical depth of the cloud, respectively. Assuming the source function is constant and the line profile is only subject to Doppler broadening, the optical depth $\tau$ can be expressed as a Gaussian profile: $\tau=\tau_{0}\text{exp}[-(\lambda-\lambda_{0}-\lambda_{0}v/c)^{2}/W^2]$, where $\tau_{0}$ is optical depth at the line center, $v$ and $W$ are the Doppler velocity and width, respectively. Applying this method to the H$\alpha$ profiles for the filament spine, we derive the optical depth $\tau_{0}$, which is plotted in Figure \ref{fig2}f. Moreover, we estimate the area of the filament, the boundary of which is determined by a threshold, i.e., 1.1 times the minimal intensity of the filament region, through the trial-and-error method. Obviously, the filament area increased almost synchronously with the optical depth $\tau_{0}$. These results clearly confirm that the filament materials were also accumulated gradually in the filament channel during its formation.

\subsection{Persistent Drainage of Filament Plasma} \label{subsec: sub2}
\begin{figure*}[ht!]
	\centering
	\includegraphics[width=0.9\textwidth,height=0.9\textwidth]{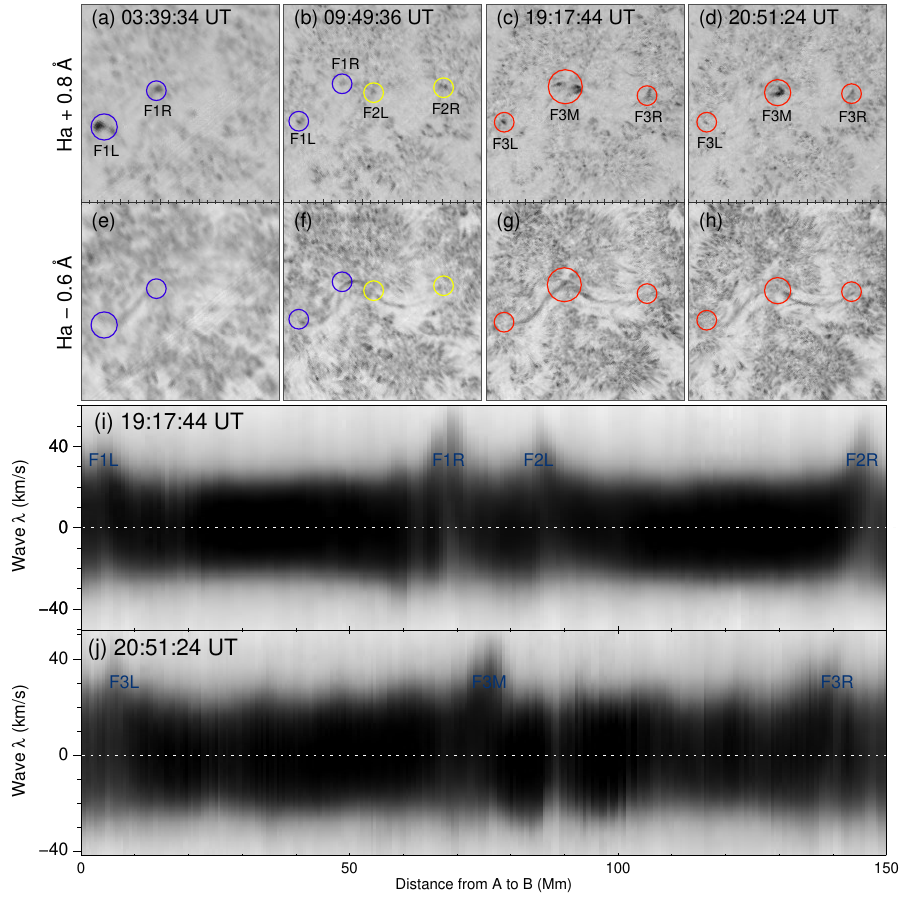}
	\caption{(a)--(d) H$\alpha$ red-wing images showing the drainage of the filament materials at the footpoints. The blue, yellow and red circles represent the footpoints of F1, F2 and F3, respectively. (e)--(h) H$\alpha$ blue-wing images at the same time as panels (a)--(d). (i)--(j) H$\alpha$ spectrum-distance diagrams for the slit AB showing the Doppler redshifts at the footpoints. The dashed lines indicate the H$\alpha$ line center. The animation that starts at 00:17 UT and ends at 22:51 is available online to show the evolution of the redshifts at all footpoints with a duration of 28 s. \label{fig3}}
\end{figure*}

A very interesting feature is that H$\alpha$ redshifts appeared at the footpoints of F1, F2, and F3 for most of the time during the filament formation. In the H$\alpha$ red-wing images (Figure \ref{fig3}a--\ref{fig3}d and attached animation), one can see that absorption prominently appeared at the two footpoints of F1 (i.e., F1L and F1R in Figure \ref{fig3}a). It was also observed at the footpoints of F2 (i.e., F2L and F2R in Figure \ref{fig3}b) when it appeared. In order to show the redshift more clearly, we select a curved slice along the filament as indicated in Figure \ref{fig2}d. The corresponding distance-spectrum diagram is displayed in Figure \ref{fig3}i, in which the H$\alpha$ spectra at the four footpoints are significantly shifted toward the red wing. It is found that the two red-shifted footpoints in the middle even merged together after $\sim$19:20 UT, as indicated by F3M in Figure \ref{fig3}c--\ref{fig3}d and Figure \ref{fig3}i--\ref{fig3}j. 

\begin{figure*}[ht!]
	\centering
	\includegraphics[width=0.9\textwidth,height=0.3\textwidth]{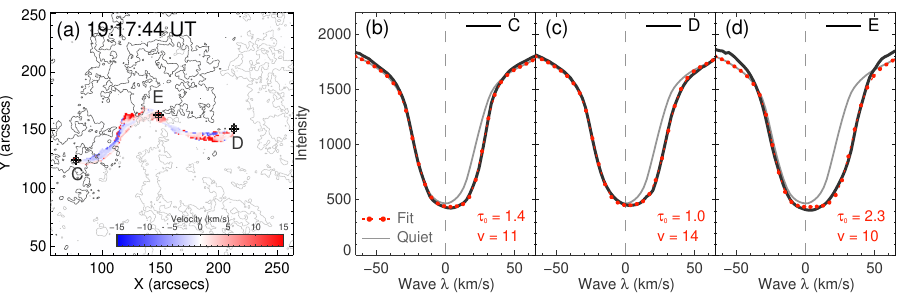}
	\caption{(a) Dopplergram of the filament F3. The grey and black contours represent the positive and negative polarities of $\pm$50G. (b)--(d) H$\alpha$ profiles (black) at points C, D, E in panel (a). The gray lines display the H$\alpha$ spectrum of the quiescent region. The dashed lines in red show the fitted results.  
      \label{fig4}}
\end{figure*}

We also apply the cloud model to all H$\alpha$ spectra at the entire filament region. The resulting Dopplergram is shown in Figure \ref{fig4}a, which clearly shows the appearance of downflows at the two footpoints (F3L and F3R) of F3. Figures \ref{fig4}b--\ref{fig4}c show two representative H$\alpha$ profiles. The fitting results give the corresponding red-shift velocities being more than 10 km s$^{-1}$. These results show that, during the formation of F3, the filament materials also drained down continuously to the footpoints, which may partly counteract the accumulation of materials within the filament channel as revealed in Section \ref{subsec: sub1}.

Besides, it is interesting to see that the H$\alpha$ profiles also present redshifts with a velocity of around 10 km s$^{-1}$ at the merging site F3M between F1 and F2. This even persisted for a long time after F3 has been well formed visually. Comparing with F1R and F2L, the redshifts at F3m are stronger as shown in Figures \ref{fig3}c--\ref{fig3}d. It is interpreted that the reconnection may take place between two dark threads of F1 and F2, thus enhancing the drainage of filament materials. In addition, during the formation of F3, the drainage at the footpoints of the fluxes of F1 and F2 that have not reconnected yet likely also has a contribution to the redshifts at F3M. 
By contrast, in the H$\alpha$ blue-wing images, the absorption feature was also detectable but primarily on the spines of F1, F2, and F3 (Figures \ref{fig3}e--\ref{fig3}h). This may be an indication of the filament materials moving out of the corresponding magnetic dips before falling down.

\subsection{Plasma Injection by EUV Jets} \label{subsec: sub3}
\begin{figure*}[ht!]
	\centering
	\includegraphics[width=0.9\textwidth,height=0.6\textwidth]{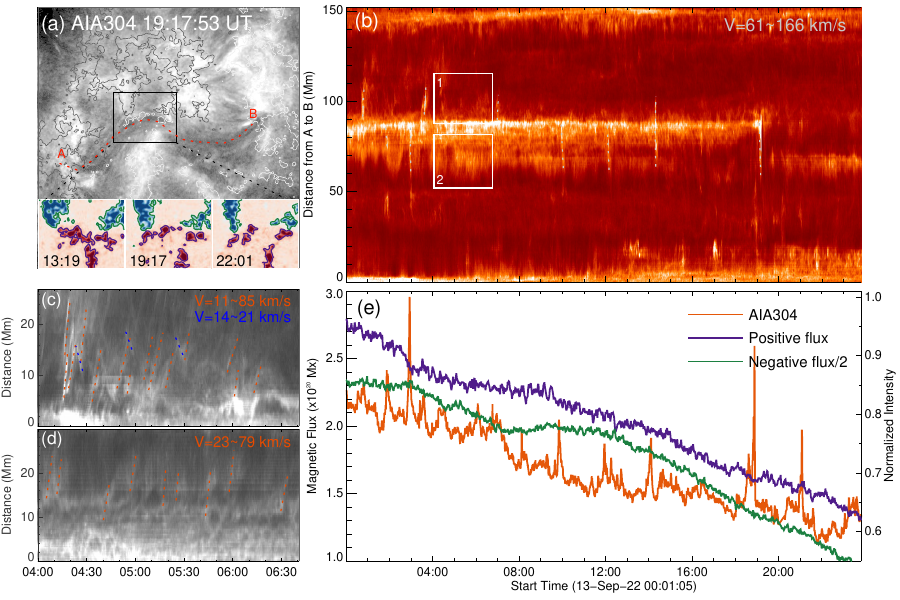}
	\caption{(a) AIA 304 {\AA} image overlaid by the contours ($\pm$50 G) of the HMI line-of-sight magnetogram. The inserted panels are the zoom-in of the magnetograms for the region of interest with the contours in blue (green) representing the positive (negative) polarities of 50 G (--50 G). (b) Distance-time diagram of the AIA 304 {\AA} images along the slit AB in panel (a). The dashed lines in white show the trajectories of the ejected jets. (c)--(d) The zoom-in of the time-distance diagram for the white boxes 1 and 2 in panel (b). The dashed red/blue lines indicate the trajectories of the upward/downward plasma. (e) Temporal evolutions of the integrated fluxes for the positive (purple) and negative (green) polarities, as well as the integrated AIA 304 {\AA} intensity (red) as shown in the black box in panel (a).   \label{fig5}}
\end{figure*}

The formation of the magnetic configuration of F3 and its material circulation are argued to be closely related to continuous outflow jets occurring near the joint region of filaments F1 and F2. These jets can be seen most obviously at the AIA 304 {\AA} passband, indicating that they have a temperature similar to the transition region and are thus hotter than the filament but cooler than the corona. To see it more clearly, we take a curved slice AB (Figure \ref{fig5}a) along the spine of F3 to create a time-distance diagram of AIA 304 {\AA} images (Figure \ref{fig5}b). It clearly shows that, at the joint region of F1 and F2, the emission was remarkably enhanced. Moreover, from the bright joint region, some prominent transition-region jets were quickly ejected outward with a velocity of 60--170 km s$^{-1}$. Then, they moved along the filament spine either toward the east or the west (Figure \ref{fig5}b). In addition, some weak upward jets were also observed. Figures \ref{fig5}c and \ref{fig5}d show the zoom-in of boxes 1 and 2 in Figure \ref{fig5}b, respectively. It is found that the weak jets, resembling spicules, were continuously injected from their bright base to the filament but with relatively small velocities ranging from 10 to 85 km s$^{-1}$. It is argued that these ejected hot plasma likely provide a source of the cold materials within the filament channel through the condensation process as suggested by \citet{Xiachun.2012.ApJL} and \citet{Xiachun.2016.ApJ}. However, it seems that not all erupted hot plasmas were supplied to the filament because part of them were observed to return to the base of the jets (see Figure \ref{fig5}c). 

The merging of the magnetic structures of F1 and F2 and the transition-region jets could be driven by flux cancellation. In the sub-panels of Figure \ref{fig5}a, we plot the HMI line-of-sight magnetograms at three instants, which show an obvious flux cancellation occurring at the joint region of F1 and F2. We further integrate the positive and negative magnetic fluxes within the region of interest and plot their temporal evolutions and find that both of them decrease constantly during the entire formation process of F3. Moreover, the time variation of the average AIA 304 {\AA} intensity within the same field-of-view is over-plotted for a comparison (Figure \ref{fig5}e). The big spikes roughly correspond to the decreases of the magnetic fluxes, in particular of the positive one. It is suggested that magnetic reconnection should play a crucial role not only in reconnecting magnetic structures of F1 and F2 to form that of F3 but also in injecting the transition-region plasma to the channel of F3, which then condense to cold materials of the filament.

\section{Summary and Discussions} \label{sec: sum}
In this paper, we present a detailed study of the formation of a long filament via merging two filament segments with the imaging and spectroscopic observations from CHASE and SDO. The H$\alpha$ line at the filament spine went through an obviously enhanced absorption and moderate broadening, suggesting the accumulation of cold materials. Meanwhile, the H$\alpha$ line at the filament foopoints presented prominent redshifts for most of the time with a downward Doppler velocity more than 10 km s$^{-1}$, indicating the drainage of filament materials. Moreover, abundant transition-region plasmas were continuously injected into the filament channel through intermittent jets. The observations document that the filament formation process involves a highly dynamic circulation of plasma that includes injection, accumulation, and drainage of filament materials simultaneously. 
  
\begin{figure*}[ht!]
	\centering
	\includegraphics[width=0.9\textwidth,height=0.432\textwidth]{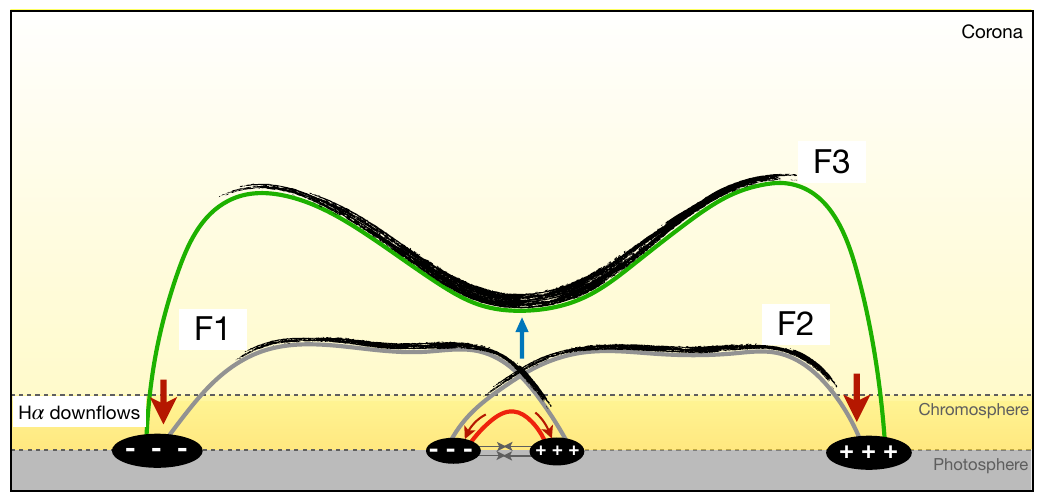}
	\caption{Schematic drawing of the formation of the long filament F3. The lines with different colors indicate magnetic field lines, which indicate magnetic flux rope structures of F1, F2 and F3. The blue arrow indicates the upward EUV jets caused by the reconnection between two threads of F1 and F2. The black threads represent the cold filament materials. The arrows in red represent downward-moving filament materials shown as H$\alpha$ redshifts.    \label{fig6}}
\end{figure*}

The circulation of hot and cold materials is believed to be an important process for the filament formation as well reproduced in previous observations and simulations \citep[e.g.,][]{Liuwei.2012.ApJL,Luna.2012.ApJ, Xiachun.2016.ApJ, Huangchujie.2021.ApJL}. For the current event, the heating and injection of the chromospheric materials into the filament are supposed to be through magnetic reconnection. As indicated by the flux cancellation occurring near the joint region of the two filament segments, the reconnection is argued to take place between the adjacent legs of F1 and F2 flux ropes. As illustrated in Figure \ref{fig6}, the right leg of F1 and the left leg of F2 approach each other, as driven by the converging motions, and reconnect over the polarity inversion line, forming a longer filament channel F3, with the submergence of the reconnected short loop manifesting as the flux cancellation \citep{Chengxin.2015.ApJ}. 

Besides connecting the magnetic structures of F1 and F2, the reconnection also heats the chromospheric plasma, which is then injected to the newly formed filament channel F3, as evidenced by the continuous AIA 304 {\AA} outflow jets. Once thermal instability occurs, the hot plasma within the filament channel starts to condense and gradually form the cold filament materials. It is worthy of mentioning that, after the formation of F3, we can not exclude the other possibility that the local hot coronal plasma directly condenses into the filament materials. As the longer magnetic field lines are created, on the one hand, the condensation will become easier as suggested by the simulations of \cite{Kaneko2017ApJ}, on the other hand, the dips become deeper to allow more cold materials to be accumulated. In case the newly formed lines without cold materials, the condensation will produce new cold materials; in case these lines already with cold materials, the condensation will continue to accumulate the filament materials. At the same time, the filament materials also drained out of their magnetic dips with the descending motions, thus giving rise to the redshifts at the footpoints of F3.

Our observations suggest that magnetic reconnection not only forms magnetic configuration of the filament channel but also supplies filament materials mainly through an injection-condensation process, which is comparable with the conclusion of \cite{Yangbo.2021.ApJL}. However, the hot plasma in their work is evaporated into the filament channel during the flare, differing from the direct injection by a series of outflow jets as found in the current study. Moreover, the filament formation in \cite{Yangbo.2021.ApJL} only lasted for tens of minutes, which is much shorter than the duration of the event we study here (about 20 hours). Such a significant difference in lifetime could be due to the fact that magnetic reconnection in \cite{Yangbo.2021.ApJL} is highly efficient and involves abundant magnetic flux in a short time, while it is relatively moderate and of small-scale in the current event and thus a series of analogous individual jets are needed \citep[e.g.,][]{Chengxin.2023.NC}. In addition, we also observe the drainage of the filament materials, which also has a role in prolonging the timescale of the filament formation.

The magnetic dips are the key for the condensation of hot plasma within the filament channel. In previous observations, magnetic dips have been observed explicitly to be formed in the high corona through the reconnection between large-scale open and closed coronal loops \citep{Lileping.2018.ApJL, Lileping.2021.ApJ, Chenhechao.2022.AA}. In contrast, in the current event, the reconnection probably occurs in the lower atmosphere because it seems that the observed jets are composed of plasma from the transition zone.

\begin{acknowledgements}
We appreciate the referee for his/her comments and suggestions. We thank the CHASE, SDO/AIA, and SDO/HMI teams for providing the high-quality data. CHASE mission is supported by China National Space Administration. The project was supported by the National Key R\&D Program of China under grants 2021YFA1600504 and by NSFC under grants 12127901 and 12333009.
\end{acknowledgements}

\bibliographystyle{aasjournal}
\bibliography{20220913}{}  

\begin{thebibliography}{}
\expandafter\ifx\csname natexlab\endcsname\relax\def\natexlab#1{#1}\fi
\providecommand{\url}[1]{\href{#1}{#1}}
\providecommand{\dodoi}[1]{doi:~\href{http://doi.org/#1}{\nolinkurl{#1}}}
\providecommand{\doeprint}[1]{\href{http://ascl.net/#1}{\nolinkurl{http://ascl.net/#1}}}
\providecommand{\doarXiv}[1]{\href{https://arxiv.org/abs/#1}{\nolinkurl{https://arxiv.org/abs/#1}}}

\bibitem[{{Antiochos} {et~al.}(1994){Antiochos}, {Dahlburg}, \&
  {Klimchuk}}]{Antiochos.1994.ApJL}
{Antiochos}, S.~K., {Dahlburg}, R.~B., \& {Klimchuk}, J.~A. 1994, \apjl, 420,
  L41, \dodoi{10.1086/187158}

\bibitem[{{Antiochos} {et~al.}(1999){Antiochos}, {MacNeice}, {Spicer}, \&
  {Klimchuk}}]{Antiochos.1999.ApJ}
{Antiochos}, S.~K., {MacNeice}, P.~J., {Spicer}, D.~S., \& {Klimchuk}, J.~A.
  1999, \apj, 512, 985, \dodoi{10.1086/306804}

\bibitem[{{Aulanier} {et~al.}(1998){Aulanier}, {Demoulin}, {van
  Driel-Gesztelyi}, {Mein}, \& {Deforest}}]{Aulanier.1998.AA}
{Aulanier}, G., {Demoulin}, P., {van Driel-Gesztelyi}, L., {Mein}, P., \&
  {Deforest}, C. 1998, \aap, 335, 309

\bibitem[{{Beckers}(1964)}]{Beckers.1964.PhDT}
{Beckers}, J.~M. 1964, PhD thesis, National Solar Observatory, Sunspot New
  Mexico

\bibitem[{{Chen} {et~al.}(2022){Chen}, {Tian}, {Li}, {Peter}, {Chitta}, \&
  {Hou}}]{Chenhechao.2022.AA}
{Chen}, H., {Tian}, H., {Li}, L., {et~al.} 2022, \aap, 659, A107,
  \dodoi{10.1051/0004-6361/202142093}

\bibitem[{{Chen} {et~al.}(2016){Chen}, {Zhang}, {Li}, \&
  {Ma}}]{Chenhuadong.2016.ApJL}
{Chen}, H., {Zhang}, J., {Li}, L., \& {Ma}, S. 2016, \apjl, 818, L27,
  \dodoi{10.3847/2041-8205/818/2/L27}

\bibitem[{{Chen} {et~al.}(2020){Chen}, {Xu}, \& {Ding}}]{Chenpengfei.2020.RAA}
{Chen}, P.-F., {Xu}, A.-A., \& {Ding}, M.-D. 2020, Research in Astronomy and
  Astrophysics, 20, 166, \dodoi{10.1088/1674-4527/20/10/166}

\bibitem[{{Cheng} {et~al.}(2015){Cheng}, {Ding}, \& {Fang}}]{Chengxin.2015.ApJ}
{Cheng}, X., {Ding}, M.~D., \& {Fang}, C. 2015, \apj, 804, 82,
  \dodoi{10.1088/0004-637X/804/2/82}

\bibitem[{{Cheng} {et~al.}(2014){Cheng}, {Ding}, {Zhang}, {Srivastava}, {Guo},
  {Chen}, \& {Sun}}]{Chengxin.2014.ApJL}
{Cheng}, X., {Ding}, M.~D., {Zhang}, J., {et~al.} 2014, \apjl, 789, L35,
  \dodoi{10.1088/2041-8205/789/2/L35}

\bibitem[{{Cheng} {et~al.}(2023){Cheng}, {Priest}, {Li}, {Chen}, {Aulanier},
  {Chitta}, {Wang}, {Peter}, {Zhu}, {Xing}, {Ding}, {Solanki}, {Berghmans},
  {Teriaca}, {Aznar Cuadrado}, {Zhukov}, {Guo}, {Long}, {Harra}, {Smith},
  {Rodriguez}, {Verbeeck}, {Barczynski}, \& {Parenti}}]{Chengxin.2023.NC}
{Cheng}, X., {Priest}, E.~R., {Li}, H.~T., {et~al.} 2023, Nature
  Communications, 14, 2107, \dodoi{10.1038/s41467-023-37888-w}

\bibitem[{{Cheung} \& {Isobe}(2014)}]{Cheung2014LRSP}
{Cheung}, M. C.~M., \& {Isobe}, H. 2014, Living Reviews in Solar Physics, 11,
  3, \dodoi{10.12942/lrsp-2014-3}

\bibitem[{Fang {et~al.}(2023)Fang, Zhang, Bi, \& Song}]{Fangyue.2023.ApJ}
Fang, Y., Zhang, J., Bi, Y., \& Song, Z. 2023, The Astrophysical Journal, 955,
  87, \dodoi{10.3847/1538-4357/acf19e}

\bibitem[{{Hong} {et~al.}(2022){Hong}, {Qiu}, {Hao}, {Xu}, {Li}, {Ding}, \&
  {Fang}}]{Hongjie.2022.AA}
{Hong}, J., {Qiu}, Y., {Hao}, Q., {et~al.} 2022, \aap, 668, A9,
  \dodoi{10.1051/0004-6361/202244427}

\bibitem[{{Huang} {et~al.}(2021){Huang}, {Guo}, {Ni}, {Xu}, \&
  {Chen}}]{Huangchujie.2021.ApJL}
{Huang}, C.~J., {Guo}, J.~H., {Ni}, Y.~W., {Xu}, A.~A., \& {Chen}, P.~F. 2021,
  \apjl, 913, L8, \dodoi{10.3847/2041-8213/abfbe0}

\bibitem[{{Jiang} {et~al.}(2014){Jiang}, {Yang}, {Wang}, {Ji}, {Liu}, {Li}, \&
  {Li}}]{Jiangyunchun.2014.ApJ}
{Jiang}, Y., {Yang}, J., {Wang}, H., {et~al.} 2014, \apj, 793, 14,
  \dodoi{10.1088/0004-637X/793/1/14}

\bibitem[{{Joshi} {et~al.}(2016){Joshi}, {Filippov}, {Schmieder}, {Magara},
  {Moon}, \& {Uddin}}]{Joshi.2016.ApJ}
{Joshi}, N.~C., {Filippov}, B., {Schmieder}, B., {et~al.} 2016, \apj, 825, 123,
  \dodoi{10.3847/0004-637X/825/2/123}

\bibitem[{{Joshi} {et~al.}(2014){Joshi}, {Magara}, \& {Inoue}}]{Joshi.2014.ApJ}
{Joshi}, N.~C., {Magara}, T., \& {Inoue}, S. 2014, \apj, 795, 4,
  \dodoi{10.1088/0004-637X/795/1/4}

\bibitem[{{Kaneko} \& {Yokoyama}(2017)}]{Kaneko2017ApJ}
{Kaneko}, T., \& {Yokoyama}, T. 2017, \apj, 845, 12,
  \dodoi{10.3847/1538-4357/aa7d59}

\bibitem[{{Lemen} {et~al.}(2012){Lemen}, {Title}, {Akin}, {Boerner}, {Chou},
  {Drake}, {Duncan}, {Edwards}, {Friedlaender}, {Heyman}, {Hurlburt}, {Katz},
  {Kushner}, {Levay}, {Lindgren}, {Mathur}, {McFeaters}, {Mitchell}, {Rehse},
  {Schrijver}, {Springer}, {Stern}, {Tarbell}, {Wuelser}, {Wolfson}, {Yanari},
  {Bookbinder}, {Cheimets}, {Caldwell}, {Deluca}, {Gates}, {Golub}, {Park},
  {Podgorski}, {Bush}, {Scherrer}, {Gummin}, {Smith}, {Auker}, {Jerram},
  {Pool}, {Soufli}, {Windt}, {Beardsley}, {Clapp}, {Lang}, \&
  {Waltham}}]{Lemen.2012.Solphys}
{Lemen}, J.~R., {Title}, A.~M., {Akin}, D.~J., {et~al.} 2012, \solphys, 275,
  17, \dodoi{10.1007/s11207-011-9776-8}

\bibitem[{{Li} {et~al.}(2019){Li}, {Fang}, {Li}, {Ding}, {Chen}, {Chen}, {Lin},
  {Chen}, {Chen}, {Tao}, {You}, {Hao}, {Dai}, {Cheng}, {Guo}, {Hong}, {An},
  {Cheng}, {Chen}, {Wang}, \& {Zhang}}]{Lichuan.2019.RAA}
{Li}, C., {Fang}, C., {Li}, Z., {et~al.} 2019, Research in Astronomy and
  Astrophysics, 19, 165, \dodoi{10.1088/1674-4527/19/11/165}

\bibitem[{{Li} {et~al.}(2022){Li}, {Fang}, {Li}, {Ding}, {Chen}, {Qiu}, {You},
  {Yuan}, {An}, {Tao}, {Li}, {Chen}, {Liu}, {Mei}, {Yang}, {Zhang}, {Cheng},
  {Chen}, {Chen}, {Gu}, {Huang}, {Liu}, {Han}, {Xin}, {Chen}, {Ni}, {Wang},
  {Rao}, {Li}, {Lu}, {Wang}, {Lin}, {Jiang}, {Meng}, \&
  {Zhao}}]{Lichuan.2022.SciCH}
---. 2022, Science China Physics, Mechanics, and Astronomy, 65, 289602,
  \dodoi{10.1007/s11433-022-1893-3}

\bibitem[{{Li} {et~al.}(2021){Li}, {Peter}, {Chitta}, \&
  {Song}}]{Lileping.2021.ApJ}
{Li}, L., {Peter}, H., {Chitta}, L.~P., \& {Song}, H. 2021, \apj, 910, 82,
  \dodoi{10.3847/1538-4357/abe537}

\bibitem[{{Li} {et~al.}(2018){Li}, {Zhang}, {Peter}, {Chitta}, {Su}, {Xia},
  {Song}, \& {Hou}}]{Lileping.2018.ApJL}
{Li}, L., {Zhang}, J., {Peter}, H., {et~al.} 2018, \apjl, 864, L4,
  \dodoi{10.3847/2041-8213/aad90a}

\bibitem[{{Liu} {et~al.}(2022){Liu}, {Tao}, {Chen}, {Han}, {Chen}, {Mei},
  {Yang}, {Hu}, {Xin}, {Li}, {Guan}, {Xue}, {Zhu}, {Hu}, {Ha}, {He}, {Fang},
  {Li}, \& {Li}}]{Liuqiang.2022.SciCH}
{Liu}, Q., {Tao}, H., {Chen}, C., {et~al.} 2022, Science China Physics,
  Mechanics, and Astronomy, 65, 289605, \dodoi{10.1007/s11433-022-1917-1}

\bibitem[{{Liu} {et~al.}(2012){Liu}, {Berger}, \& {Low}}]{Liuwei.2012.ApJL}
{Liu}, W., {Berger}, T.~E., \& {Low}, B.~C. 2012, \apjl, 745, L21,
  \dodoi{10.1088/2041-8205/745/2/L21}

\bibitem[{{Luna} {et~al.}(2012){Luna}, {Karpen}, \& {DeVore}}]{Luna.2012.ApJ}
{Luna}, M., {Karpen}, J.~T., \& {DeVore}, C.~R. 2012, \apj, 746, 30,
  \dodoi{10.1088/0004-637X/746/1/30}

\bibitem[{{Mackay} {et~al.}(2010){Mackay}, {Karpen}, {Ballester}, {Schmieder},
  \& {Aulanier}}]{Mackay.2010.SSR}
{Mackay}, D.~H., {Karpen}, J.~T., {Ballester}, J.~L., {Schmieder}, B., \&
  {Aulanier}, G. 2010, \ssr, 151, 333, \dodoi{10.1007/s11214-010-9628-0}

\bibitem[{{Manchester} {et~al.}(2004){Manchester}, {Gombosi}, {DeZeeuw}, \&
  {Fan}}]{Manchester.2004.ApJ}
{Manchester}, W., I., {Gombosi}, T., {DeZeeuw}, D., \& {Fan}, Y. 2004, \apj,
  610, 588, \dodoi{10.1086/421516}

\bibitem[{{Martin}(1998)}]{Martin.1998.Solphys}
{Martin}, S.~F. 1998, \solphys, 182, 107, \dodoi{10.1023/A:1005026814076}

\bibitem[{{Moore} {et~al.}(2001){Moore}, {Sterling}, {Hudson}, \&
  {Lemen}}]{Moore.2001.ApJ}
{Moore}, R.~L., {Sterling}, A.~C., {Hudson}, H.~S., \& {Lemen}, J.~R. 2001,
  \apj, 552, 833, \dodoi{10.1086/320559}

\bibitem[{{Okamoto} {et~al.}(2008){Okamoto}, {Tsuneta}, {Lites}, {Kubo},
  {Yokoyama}, {Berger}, {Ichimoto}, {Katsukawa}, {Nagata}, {Shibata},
  {Shimizu}, {Shine}, {Suematsu}, {Tarbell}, \& {Title}}]{Okamoto2008ApJ}
{Okamoto}, T.~J., {Tsuneta}, S., {Lites}, B.~W., {et~al.} 2008, \apjl, 673,
  L215, \dodoi{10.1086/528792}

\bibitem[{{Ouyang} {et~al.}(2017){Ouyang}, {Zhou}, {Chen}, \&
  {Fang}}]{Ouyangyu.2017.ApJ}
{Ouyang}, Y., {Zhou}, Y.~H., {Chen}, P.~F., \& {Fang}, C. 2017, \apj, 835, 94,
  \dodoi{10.3847/1538-4357/835/1/94}

\bibitem[{{Patsourakos} {et~al.}(2020){Patsourakos}, {Vourlidas},
  {T{\"o}r{\"o}k}, {Kliem}, {Antiochos}, {Archontis}, {Aulanier}, {Cheng},
  {Chintzoglou}, {Georgoulis}, {Green}, {Leake}, {Moore}, {Nindos}, {Syntelis},
  {Yardley}, {Yurchyshyn}, \& {Zhang}}]{Patsourakos.2020.SSR}
{Patsourakos}, S., {Vourlidas}, A., {T{\"o}r{\"o}k}, T., {et~al.} 2020, \ssr,
  216, 131, \dodoi{10.1007/s11214-020-00757-9}

\bibitem[{{Pesnell} {et~al.}(2012){Pesnell}, {Thompson}, \&
  {Chamberlin}}]{Pesnell.2012.Solphys}
{Pesnell}, W.~D., {Thompson}, B.~J., \& {Chamberlin}, P.~C. 2012, \solphys,
  275, 3, \dodoi{10.1007/s11207-011-9841-3}

\bibitem[{{Qiu} {et~al.}(2022){Qiu}, {Rao}, {Li}, {Fang}, {Ding}, {Li}, {Ni},
  {Wang}, {Hong}, {Hao}, {Dai}, {Chen}, {Wan}, {Xu}, {You}, {Yuan}, {Tao},
  {Li}, {He}, \& {Liu}}]{Qiuye.2022.SciCH}
{Qiu}, Y., {Rao}, S., {Li}, C., {et~al.} 2022, Science China Physics,
  Mechanics, and Astronomy, 65, 289603, \dodoi{10.1007/s11433-022-1900-5}

\bibitem[{{Scherrer} {et~al.}(2012){Scherrer}, {Schou}, {Bush}, {Kosovichev},
  {Bogart}, {Hoeksema}, {Liu}, {Duvall}, {Zhao}, {Title}, {Schrijver},
  {Tarbell}, \& {Tomczyk}}]{Scherrer.2012.Solphys}
{Scherrer}, P.~H., {Schou}, J., {Bush}, R.~I., {et~al.} 2012, \solphys, 275,
  207, \dodoi{10.1007/s11207-011-9834-2}

\bibitem[{{Schmieder} {et~al.}(2004){Schmieder}, {Mein}, {Deng}, {Dumitrache},
  {Malherbe}, {Staiger}, \& {Deluca}}]{Schmieder.2004.Solphys}
{Schmieder}, B., {Mein}, N., {Deng}, Y., {et~al.} 2004, \solphys, 223, 119,
  \dodoi{10.1007/s11207-004-1107-x}

\bibitem[{{Schmieder} {et~al.}(2014){Schmieder}, {Roudier}, {Mein}, {Mein},
  {Malherbe}, \& {Chandra}}]{Schmieder.2014.AA}
{Schmieder}, B., {Roudier}, T., {Mein}, N., {et~al.} 2014, \aap, 564, A104,
  \dodoi{10.1051/0004-6361/201322861}

\bibitem[{{Schou} {et~al.}(2012){Schou}, {Scherrer}, {Bush}, {Wachter},
  {Couvidat}, {Rabello-Soares}, {Bogart}, {Hoeksema}, {Liu}, {Duvall}, {Akin},
  {Allard}, {Miles}, {Rairden}, {Shine}, {Tarbell}, {Title}, {Wolfson},
  {Elmore}, {Norton}, \& {Tomczyk}}]{Schou.2012.Solphys}
{Schou}, J., {Scherrer}, P.~H., {Bush}, R.~I., {et~al.} 2012, \solphys, 275,
  229, \dodoi{10.1007/s11207-011-9842-2}

\bibitem[{{Tandberg-Hanssen}(1974)}]{Tandberg.1974.Book}
{Tandberg-Hanssen}, E. 1974, {Solar Prominences}, Vol.~12

\bibitem[{{T{\"o}r{\"o}k} {et~al.}(2011){T{\"o}r{\"o}k}, {Chandra}, {Pariat},
  {D{\'e}moulin}, {Schmieder}, {Aulanier}, {Linton}, \&
  {Mandrini}}]{Torok.2011.ApJ}
{T{\"o}r{\"o}k}, T., {Chandra}, R., {Pariat}, E., {et~al.} 2011, \apj, 728, 65,
  \dodoi{10.1088/0004-637X/728/1/65}

\bibitem[{{T{\"o}r{\"o}k} \& {Kliem}(2003)}]{Torok.2003.AA}
{T{\"o}r{\"o}k}, T., \& {Kliem}, B. 2003, \aap, 406, 1043,
  \dodoi{10.1051/0004-6361:20030692}

\bibitem[{{van Ballegooijen} \& {Martens}(1989)}]{vanBallegooijen.1989.ApJ}
{van Ballegooijen}, A.~A., \& {Martens}, P.~C.~H. 1989, \apj, 343, 971,
  \dodoi{10.1086/167766}

\bibitem[{{Wang} {et~al.}(2018){Wang}, {Yan}, {Qu}, {UeNo}, {Ichimoto}, {Deng},
  {Cao}, \& {Liu}}]{Wangjincheng.2018.ApJ}
{Wang}, J., {Yan}, X., {Qu}, Z., {et~al.} 2018, \apj, 863, 180,
  \dodoi{10.3847/1538-4357/aad187}

\bibitem[{{Wang} {et~al.}(2022){Wang}, {Yan}, {Xue}, {Yang}, {Li}, {Chen},
  {Xia}, \& {Liu}}]{Wangjincheng.2022.AA}
{Wang}, J., {Yan}, X., {Xue}, Z., {et~al.} 2022, \aap, 659, A76,
  \dodoi{10.1051/0004-6361/202142584}

\bibitem[{{Wang}(1999)}]{Wangyuming.1999.ApJL}
{Wang}, Y.~M. 1999, \apjl, 520, L71, \dodoi{10.1086/312149}

\bibitem[{{Xia} {et~al.}(2012){Xia}, {Chen}, \& {Keppens}}]{Xiachun.2012.ApJL}
{Xia}, C., {Chen}, P.~F., \& {Keppens}, R. 2012, \apjl, 748, L26,
  \dodoi{10.1088/2041-8205/748/2/L26}

\bibitem[{{Xia} \& {Keppens}(2016)}]{Xiachun.2016.ApJ}
{Xia}, C., \& {Keppens}, R. 2016, \apj, 823, 22,
  \dodoi{10.3847/0004-637X/823/1/22}

\bibitem[{{Yan} {et~al.}(2016){Yan}, {Priest}, {Guo}, {Xue}, {Wang}, \&
  {Yang}}]{Yanxiaoli.2016.ApJ}
{Yan}, X.~L., {Priest}, E.~R., {Guo}, Q.~L., {et~al.} 2016, \apj, 832, 23,
  \dodoi{10.3847/0004-637X/832/1/23}

\bibitem[{{Yan} {et~al.}(2015){Yan}, {Xue}, {Pan}, {Wang}, {Xiang}, {Kong}, \&
  {Yang}}]{Yanxiaoli.2015.ApJS}
{Yan}, X.~L., {Xue}, Z.~K., {Pan}, G.~M., {et~al.} 2015, \apjs, 219, 17,
  \dodoi{10.1088/0067-0049/219/2/17}

\bibitem[{{Yang} {et~al.}(2021){Yang}, {Yang}, {Bi}, {Hong}, \&
  {Xu}}]{Yangbo.2021.ApJL}
{Yang}, B., {Yang}, J., {Bi}, Y., {Hong}, J., \& {Xu}, Z. 2021, \apjl, 921,
  L33, \dodoi{10.3847/2041-8213/ac31b6}

\bibitem[{{Yang} {et~al.}(2017){Yang}, {Yan}, {Li}, {Xue}, \&
  {Xiang}}]{Yangliheng.2017.ApJ}
{Yang}, L., {Yan}, X., {Li}, T., {Xue}, Z., \& {Xiang}, Y. 2017, \apj, 838,
  131, \dodoi{10.3847/1538-4357/aa653a}

\bibitem[{{Zhao} {et~al.}(2017){Zhao}, {Xia}, {Keppens}, \&
  {Gan}}]{Zhaoxiaozhou.2017.ApJ}
{Zhao}, X., {Xia}, C., {Keppens}, R., \& {Gan}, W. 2017, \apj, 841, 106,
  \dodoi{10.3847/1538-4357/aa7142}

\bibitem[{{Zhou} {et~al.}(2020){Zhou}, {Chen}, {Hong}, \&
  {Fang}}]{Zhouyuhao.2020.NA}
{Zhou}, Y.~H., {Chen}, P.~F., {Hong}, J., \& {Fang}, C. 2020, Nature Astronomy,
  4, 994, \dodoi{10.1038/s41550-020-1094-3}

\bibitem[{{Zhu} {et~al.}(2015){Zhu}, {Liu}, {Alexander}, {Sun}, \&
  {McAteer}}]{Zhuchunming.2015.ApJ}
{Zhu}, C., {Liu}, R., {Alexander}, D., {Sun}, X., \& {McAteer}, R.~T.~J. 2015,
  \apj, 813, 60, \dodoi{10.1088/0004-637X/813/1/60}

\end{thebibliography}

\end{document}